# Stress channelling in extreme couple-stress materials Part II: Localized folding vs faulting of a continuum in single and cross geometries

Panos A. Gourgiotis, Davide Bigoni[1]
email: p.gourgiotis@unitn.it, bigoni@ing.unitn.it
University of Trento, via Mesiano 77, I-38123 Trento, Italy

**Abstract**. The antiplane strain Green's functions for an applied concentrated force and moment are obtained for Cosserat elastic solids with extreme anisotropy, which can be tailored to bring the material in a state close to an instability threshold such as failure of ellipticity. It is shown that the wave propagation condition (and not ellipticity) governs the behavior of the antiplane strain Green's functions. These Green's functions are used as perturbing agents to demonstrate in an extreme material the emergence of localized (single and cross) stress channelling and the emergence of antiplane localized *folding* (or creasing, or weak elastostatic shock) and *faulting* (or elastostatic shock) of a Cosserat continuum, phenomena which remain excluded for a Cauchy elastic material. During folding some components of the displacement gradient suffer a *finite* jump, whereas during faulting the displacement itself displays a *finite* discontinuity.

## 1. Introduction

The mechanical conditions leading to the formation of a periodic pattern of creases in an elastic solid under load have attracted great interest (see for instance Onuki, 1989; Roccabianca et al., 2010; Hutchinson, 2013; Jin and Suo, 2015), particularly with reference to soft matter in biological applications, for altering wettability, and for flexible electronics. It is not difficult to explain periodic creasing in terms of bifurcation theory for elastic solids, but the formation of an isolated crease, or fold (to be distinguished from *bend*[2]), is hard to obtain and perhaps even impossible for a Cauchy elastic continuum. It is shown in the present article that folding (which induces a localized and finite discontinuity in the displacement gradient) and faulting (which induces a localized and finite discontinuity in the displacement itself) can be achieved in a Cosserat elastic material at the limit of failure of ellipticity. However, the discontinuities become already visible (in terms of a rapid variation of the displacement or its gradient instead of a jump) near this limit, when ellipticity is still preserved[3]. The fact that these discontinuities are clearly visible in the proximity (but still inside) of the border of the elliptic domain (and even when the strain energy function is still strictly positive), means that extreme materials as those analyzed in the present article *can be realized in practice* and employed to explore unchallenged mechanical behavior.

---
[1] *Corresponding author*. Email: bigoni@ing.unitn.it

The perturbative approach (Bigoni, 2012) employed to show the formation of discontinuities is based on the determination of the infinite-body Green's function, which is obtained here (under antiplane deformation conditions for an anisotropic Cosserat elastic material specified in Part I of this study) in the two cases of applied concentrated force and moment (the latter being possible because the continuum involves body moments). The Green's functions are obtained under the condition that (planar) waves can propagate with a finite speed inside the body – the (WP) condition – and not under the assumption of ellipticity. This implies that the Green's function can still be used in the special case where ellipticity is lost, but waves can propagate (a situation possible for Cosserat elasticity, Part I). Folding has been found to occur in single or cross geometries (respectively at the Elliptic-Imaginary/Parabolic or at the Elliptic-Complex/Hyperbolic borders) and is found to decay exponentially in both cases (with exceptions documented for non-decaying single folding). Finally, at these two boundaries of ellipticity failure, the application of a concentrated moment has revealed an unexpected feature, namely, the single or cross *faulting* of a Cosserat continuum. Single (or cross) faulting is the emergence of one (or two) finite size discontinuity surface for displacements dividing the material into two (or four) parts. The emergence of folding and faulting revealed by the perturbative approach is also substantiated with analytical solutions (possible only in special cases) demonstrating a situation of non-decaying single folding and another situation in which a Dirac localization (possible for an extreme material possessing only a non-vanishing Cosserat bending stiffness) is found.

---

[2] Bifurcations in solids are usually identified with the occurrence of stationary waves (Biot, 1965; Hill and Hutchinson, 1975; Bigoni, 2012), which, roughly speaking, correspond to *bending*, not *folding*.
[3] This situation shares analogies with the formation of shear bands in a solid, which formally occurs at the elliptic border (Rice, 1977), but can become already visible near this border using a perturbative approach (Bigoni and Capuani, 2002, 2005; Bigoni, 2012).

From the findings documented in the present article (restricted for simplicity to antiplane strain conditions) several conclusions can be drawn from the inclusion of Cosserat effects as a remediation to pathological mesh dependence for a constitutive model violating a material instability threshold. More importantly, the fact that Cosserat extreme materials can be realized in practice[4] opens new and unexpected applicative possibilities, such as localization and channelling of signals, even in complex geometries involving isolated or repeated single or cross folding or faulting.

## 2. Antiplane Green's functions

The Green's functions for concentrated force and moment in an orthotropic Cosserat material subject to antiplane deformation are derived in this Section, by employing a Fourier transformation technique sharing analogies with the standard technique in classical Cauchy elasticity (Willis, 1971; 1973; 1991).

2.1 *The analyzed Cosserat materials*

Constrained rotation Cosserat (linear elastic) materials are considered as described in Part I, with a particular emphasis on extreme materials for which loss of ellipticity is approached, but with positive definiteness (PD) of the elasticity still preserved (thus defining a material in which waves can propagate and the solution is unique and stable). In particular, under antiplane strain conditions $(\text{PD})^{\mathbb{C}} = (\text{SE})^{\mathbb{C}}$, so that failure of these criteria occur simultaneously (and imply failure of ellipticity for a *classical* material), but $(\text{PD})^{\mathbb{B}}$ and $(\text{SE})^{\mathbb{B}}$ remain separate criteria. The material parameters will therefore be selected so that (PD), (SE) and also (E) fail simultaneously. In this way, the analyzed materials approach a material instability, but from a domain in which uniqueness and positive definiteness of strain energy hold, and waves can still propagate in the medium.

---

[4] Studies by, among others, Banks and Sokolowski (1968), Lakes (1995), Ostoja-Starzewski (1999), Bigoni and Drugan (2007) provide techniques for the determination of the Cosserat elastic moduli in isotropic and anisotropic solids.

## 2.2 Infinite-body Green's functions for concentrated force and moment

The field equation governing antiplane deformations in the case of an orthotropic Cosserat material (see also Section 7.1, Part I) admits two infinite-body Green's functions: one for an out-of-plane concentrated force $S$, and one for an in-plane concentrated moment (with components $M_x$ and $M_y$). The equilibrium equation for the out-of-plane displacement component $w(x, y)$ assumes then the following form (Eq. (91), in Part I)

$$Lw + S\delta(x)\delta(y) - \frac{1}{2}\left[M_x \delta(x)\delta'(y) - M_y \delta'(x)\delta(y)\right] = 0, \qquad (1)$$

where $\delta(\ )$ is the Dirac delta distribution and $\delta'(\ )$ denotes the derivative of the Dirac delta with respect to the pertinent variable (Gelfand and Shilov, 1964). The fourth-order differential operator $L$ is defined as

$$L = c_{55}\partial_x^2 + c_{44}\partial_y^2 - \frac{1}{4}\left(b_2 \partial_x^4 + 2b_0 \partial_x^2 \partial_y^2 + b_4 \partial_y^4\right). \qquad (2)$$

An exact solution to Eq. (1) is obtained by employing the double exponential Fourier transform. The direct and inverse double Fourier transforms are defined as

$$\tilde{w}(k_1, k_2) = \int_{-\infty}^{+\infty}\int_{-\infty}^{+\infty} w(x, y) e^{i(k_1 x + k_2 y)} dx dy, \qquad (3)$$

$$w(x, y) = \frac{1}{4\pi^2} \int_{-\infty}^{+\infty}\int_{-\infty}^{+\infty} \tilde{w}(k_1, k_2) e^{-i(k_1 x + k_2 y)} dk_1 dk_2. \qquad (4)$$

In the case of a *concentrated force*, the Green's function is derived by applying the direct double Fourier transform (3) to the field equation (1) with $M_x = M_y = 0$, yielding the out-of-plane displacement

$$w(x, y) = \frac{S}{4\pi^2} \int_{-\infty}^{\infty}\int_{-\infty}^{\infty} \frac{1}{D(k_1, k_2)} e^{-i(k_1 x + k_2 y)} dk_1 dk_2, \qquad (5)$$

where

$$D(k_1, k_2) \equiv A_{33}(k_1, k_2) = c_{55}k_1^2 + c_{44}k_2^2 + \frac{1}{4}\left(b_2 k_1^4 + 2b_0 k_1^2 k_2^2 + b_4 k_2^4\right), \tag{6}$$

is the characteristic polynomial (identified with the $A_{33}$ component of the acoustic tensor, Eq. 113, Part I) which is thus *strictly positive* when the wave propagation condition (WP) holds. Therefore, from the point of view of finding the infinite-body Green's function the (WP) condition plays the major role.

It is noted that in the classical elasticity case the characteristic polynomial and the out-of-plane displacement reduce, respectively, to (Ting, 1996)

$$D(k_1, k_2) = c_{55}k_1^2 + c_{44}k_2^2, \quad w^{cl}(x, y) = -\frac{S}{2\pi c_{44} \sqrt{\varepsilon}} \text{Log}\left[\sqrt{x^2 + \varepsilon y^2}\right], \tag{7}$$

while the non-vanishing Cauchy shear stress components become

$$\sigma_{xz} = -\frac{S\sqrt{\varepsilon}\, x}{2\pi(x^2 + \varepsilon y^2)}, \quad \sigma_{yz} = -\frac{S\sqrt{\varepsilon}\, y}{2\pi(x^2 + \varepsilon y^2)}, \tag{8}$$

where $\varepsilon = c_{55}/c_{44}$ is the ratio between the two Cauchy shear moduli.

For the evaluation of the inversion integral in Eq. (5), the integrand is factored by finding the roots of the characteristic quartic polynomial (6). The four roots can be written in the following concise way

$$k_2 = \pm i p_1(k_1), \quad k_2 = \pm i p_2(k_1), \tag{9}$$

where

$$p_{1,2} \equiv p_{1,2}(k_1) = \frac{\sqrt{b_0 k_1^2 + 2c_{44} \pm \Delta}}{\sqrt{b_4}} \quad \text{and} \quad \Delta = \sqrt{\left(b_0 k_1^2 + 2c_{44}\right)^2 - b_4 k_1^2 \left(b_2 k_1^2 + 4c_{55}\right)}. \tag{10}$$

Assuming that the (WP) condition holds, and depending on the values of the transformed variable $k_1$ ($k_1 \in \mathbb{R}$), the four roots of the characteristic polynomial in Eq. (9) can be either purely imaginary or complex conjugates, with $\text{Re}(p_{1,2}) > 0$. Accordingly, the characteristic polynomial can be written as

$$D(k_1, k_2) = \frac{b_4}{4}(k_2^2 + p_1^2)(k_2^2 + p_2^2), \tag{11}$$

so that the integrand is now decomposed into four rational parts. For each part, the residue theorem is employed in conjunction with Jordan's lemma to evaluate the integral with respect to the complex variable $k_2$. The original integration path running along the real axis in the $k_2$-plane is replaced then by a closed contour $C$, which for $y > 0$ is taken in the lower $k_2$-plane with $\text{Im}(k_2) < 0$, so that the integrand is decaying as $|k_2| \to \infty$. Similarly, when $y < 0$, the closed contour $C$ is taken in the upper $k_2$-plane, so that the following result can be derived

$$\oint_C \frac{e^{-ik_2 y}}{D(k_1, k_2)} dk_2 = \frac{2\pi \left(p_1 e^{-p_2|y|} - p_2 e^{-p_1|y|}\right)}{p_1 p_2 \Delta}, \tag{12}$$

which is valid for all $y$. Further, by noticing that $p_1(k_1)$ and $p_2(k_1)$ are even functions of their argument and by taking also into account Eq. (12), the integral in Eq. (5) becomes

$$w(x, y) = \frac{S}{\pi} \int_0^\infty Q_S(k_1, y) \cos(k_1 x) dk_1, \tag{13}$$

where

$$Q_S(k_1, y) = \frac{\left(p_1 e^{-p_2|y|} - p_2 e^{-p_1|y|}\right)}{p_1 p_2 \Delta}. \tag{14}$$

The inversion integral in Eq. (13) is divergent since $Q_S(k_1, y) = O(k_1^{-1})$ as $k_1 \to 0$, and, thus, has to be interpreted in the finite part sense (a situation analogous to the classical elasticity case). To this purpose, it is expedient to decompose the out-of-plane displacement into three integrals, in the following way

$$w(x,y) = I_{cl}(x,y) + I_{cs_1}(x,y) + I_{cs_2}(x,y),$$  (15)

where

$$I_{cl}(x,y) = \frac{S}{2\pi\sqrt{c_{44}c_{55}}} \text{F.P.} \int_0^\infty \frac{1}{k_1} e^{-|y|\sqrt{\varepsilon}k_1} \cos(k_1 x) dk_1,$$  (16)

$$I_{cs_1}(x,y) = -\frac{S\ell}{2\pi\sqrt{c_{44}c_{55}}} \int_0^\infty \frac{e^{-|y|\frac{\sqrt{\varepsilon}}{\ell}\sqrt{1+\ell^2 k_1^2}}}{\sqrt{1+\ell^2 k_1^2}} \cos(k_1 x) dk_1,$$  (17)

$$I_{cs_2}(x,y) = \frac{S}{\pi} \int_0^\infty \hat{Q}_S(k_1,y) \cos(k_1 x) dk_1,$$  (18)

in which

$$\hat{Q}_S(k_1,y) = Q_S(k_1,y) - \frac{1}{2c_{44}\sqrt{\varepsilon}} \left( \frac{e^{-|y|\sqrt{\varepsilon}k_1}}{k_1} - \frac{\ell e^{-|y|\frac{\sqrt{\varepsilon}}{\ell}\sqrt{1+\ell^2 k_1^2}}}{\sqrt{1+\ell^2 k_1^2}} \right),$$  (19)

with $\ell$ being a characteristic material length, defined through the relation $b_4 = 4c_{44}\ell^2$, which is related with the bending stiffness in the *x*-direction. The integrals in Eqs. (16) and (17) can be evaluated analytically (Gradshteyn and Ryzhik, 1980). In particular, the divergent integral in Eq. (16) corresponds to the classical elasticity solution given by Eq. (7)$_2$ and is evaluated in the finite part sense (F.P.). On the other hand, the integral in Eq. (18) is uniformly convergent and can be evaluated numerically taking into account its oscillatory character. Accordingly, the out-of-plane displacement becomes

$$w(x,y) = -\frac{S}{2\pi\sqrt{c_{44}c_{55}}} \text{Log}[z] - \frac{S}{2\pi\sqrt{c_{44}c_{55}}} K_0[z] + I_{cs_2}(x,y),$$  (20)

where $z = \ell^{-1}\sqrt{x^2 + \varepsilon y^2}$, and $K_0[\ ]$ is the modified Bessel function of the second-kind and zero order. By noting that $K_0[z] = -O(\text{Log}[z])$ as $z \to 0$ and that the function $I_{cs_2}(x,y)$ is regular at the point of application of the concentrated force, it can immediately be inferred that *the classical logarithmic singularity is eliminated* by the Cosserat effect. It is worth noting that in the isotropic case the inversion of the integral in Eq. (5) can be performed analytically (as suggested by e.g. Nowacki, 1984) yielding

$$w(x,y) = -\frac{S}{2\pi\mu}\left(K_0[r/\ell'] + \text{Log}[r/\ell']\right), \tag{21}$$

where $r = \sqrt{x^2 + y^2}$ is the distance from the origin, and $\ell' = \sqrt{\eta/\mu}$ is the pertinent characteristic length for an isotropic couple-stress material (Mindlin, 1963).

In light of the above, it can be concluded that *the couple-stress solution predicts a bounded and continuous displacement at the point of application of the load* and, therefore, 'corrects' (in a boundary-layer sense) the classical elasticity solution, which predicts a logarithmically singular behavior at the source point for an orthotropic material, Eq. (7). This finding is also in contrast with the respective result of the standard micropolar elasticity theory where the out-of-plane displacement remains logarithmically singular at the origin (Dyszlewicz, 2012). Note that under plane strain conditions the displacements at the application point of a concentrated force would also logarithmically diverge in the context of constrained Cosserat elasticity, but the investigation of this problem falls outside the scope of this study and will be addressed elsewhere. Furthermore, it is remarked that as $r \to \infty$, the couple-stress solution becomes logarithmically unbounded as in the classical theory, since away from the origin the couple-stress effects decay and the classical elasticity solution dominates. Such a pathological behavior is known in 2D elastostatic problems involving concentrated loads in the context of classical elasticity (Turteltaub and Sternberg, 1968) but also in the context of higher-order gradient elasticity theories (Georgiadis and Anagnostou, 2008).

Finally, note that the (asymmetric) shear stresses for the concentrated force case can be derived by direct substitution of the displacement solution (13) into Eqs. (87) and (88) in Part I. The explicit formulas are provided in Appendix A.

In the case of a *concentrated moment*, a procedure analogous to that presented before can be employed to derive the Green's function, assuming that $S = 0$. The displacement field produced by the concentrated moment can be written as

$$w(x,y) = \frac{M_x}{2\pi}\int_0^\infty Q_{M_x}(k_1,y)\cos(k_1 x)dk_1 + \frac{M_y}{2\pi}\int_0^\infty Q_{M_y}(k_1,y)\sin(k_1 x)dk_1, \quad (22)$$

where

$$Q_{M_x}(k_1,y) = \text{sgn}(y)\left(e^{-p_2|y|} - e^{-p_1|y|}\right)\Delta^{-1}, \quad Q_{M_y}(k_1,y) = -k_1 Q_S(k_1,y). \quad (23)$$

Note that since the functions $Q_{M_x}(k_1,y)$ and $Q_{M_y}(k_1,y)$ are bounded in $k_1 \in [0,\infty)$, the integrals in Eq. (22) are convergent and can be directly evaluated numerically taking into account their oscillatory character.

In the following, the Green's functions will be utilized as perturbing agents in several cases. Note that the employed Cosserat orthotropic material is characterized effectively by three dimensionless parameters, namely, the ratio between the shear moduli $\varepsilon = c_{55}/c_{44}$, the ratio between the bending moduli $\beta = b_2/b_4$, and the ratio $\gamma = b_0/b_4$, with $b_0 = b_1 - b_3$. In what follows, unless otherwise stated, it will be assumed that $b_4 > 0$. Finally, it is recalled that in the case of isotropy: $\varepsilon = 1$, $\beta = 1$, and $\gamma = 1$.

## 3. 'Non-extreme' behavior

Some illustrative results – relative to 'non-extreme' materials with mechanical behavior still far from instabilities – are now presented regarding the behavior of the out-of-plane displacement for the cases of a concentrated force and a concentrated moment, acting in an infinite orthotropic couple-stress medium under antiplane strain conditions.

Fig. 1 depicts the dimensionless out-of-plane displacement $wc_{44}$, plotted with respect to the dimensionless distances $x/\ell$ and $y/\ell$, for an orthotropic Cosserat material far from instability boundaries (with $\varepsilon = 1/4$, $\beta = 1/5$, and $\gamma = 1/4$) as produced by a concentrated antiplane unit force (acting at the origin of the axes). It is observed that the couple-stress solution predicts a *bounded* displacement at the point of application of the load. This is more clearly depicted in Fig. 2, where the couple stress solution is compared (along the line $y = 0$) with the classical solution, which exhibits a logarithmically unbounded behavior at the origin. It is shown that in the range $r < 2\ell$ the couple-stress effects play a significant role in the material response, while, outside this region, the couple-stress solution approaches the classical one (Fig. 2).

To examine the effects of the Cosserat anisotropy, a qualitative comparison of the out-of-plane displacement contours is presented in Fig. 3, between two materials which are *both* isotropic in the classical sense, $c_{44} = c_{55}$ ($\varepsilon = 1$), but possess different types of intrinsic Cosserat microstructure: isotropic in one case ($\beta = 1$, $\gamma = 1$), and orthotropic in the other ($\beta = 1/5$, $\gamma = 1/4$). As it is shown in Fig. 3, the Cosserat orthotropy (microstructural anisotropy) affects significantly the displacement in the vicinity of the concentrated force through a distortion of the circular contours typical of isotropy, which become concentric ellipses for orthotropy.

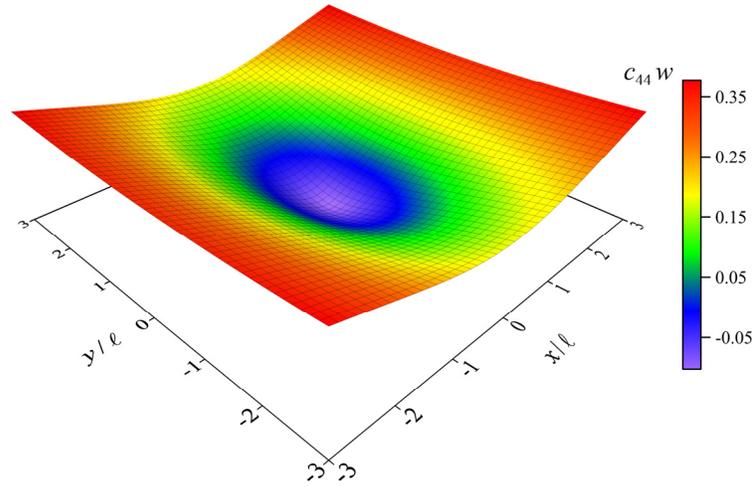

**Fig. 1**: Dimensionless out-of-plane displacement due to an antiplane concentrated unit force in an orthotropic Cosserat solid far from material instability ($\varepsilon = 1/4$, $\beta = 1/5$, $\gamma = 1/4$).

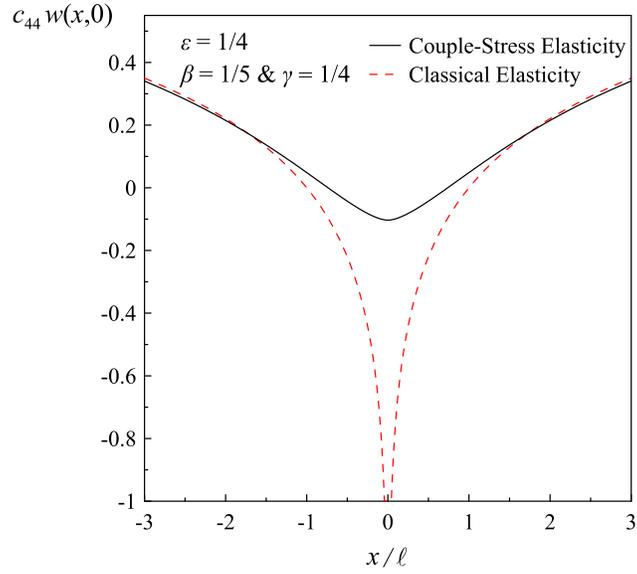

**Fig. 2**: Dimensionless out-of-plane displacement produced by an antiplane unit force: the comparison between couple-stress theory and classical elasticity at $y = 0$ reveals that the Cosserat solution (solid line) is bounded and that it approaches the Cauchy elastic solution (dashed line) as the distance from the origin increases.

The dimensionless out-of-plane displacement $w\ell c_{44}$ produced by a concentrated unit moment acting in the $y$-direction in a Cosserat solid far from material instability ($\varepsilon = 1/4$, $\beta = 1/5$, and $\gamma = 1/4$) is plotted in Fig. 4. It is worth noting that there is no counterpart of such type of loading in the classical theory of elasticity. As expected from symmetry, the displacement is zero at the point of application of the concentrated moment. Moreover, the absolute value of the displacement exhibits a bounded maximum near the origin and then decays monotonically to zero as $r \to \infty$.

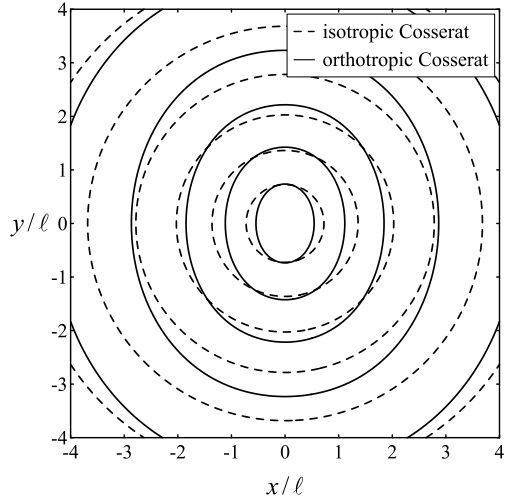

**Fig. 3**: Dimensionless level sets of the out-of-plane displacement due to an antiplane concentrated unit force for two solids: one fully isotropic ( $\varepsilon = 1$, $\beta = 1$, $\gamma = 1$ – dashed line), and the other isotropic with respect to the Cauchy part but orthotropic in the Cosserat part ( $\varepsilon = 1$, $\beta = 1/5$, $\gamma = 1/4$ – solid line). Both solids are far from material instability.

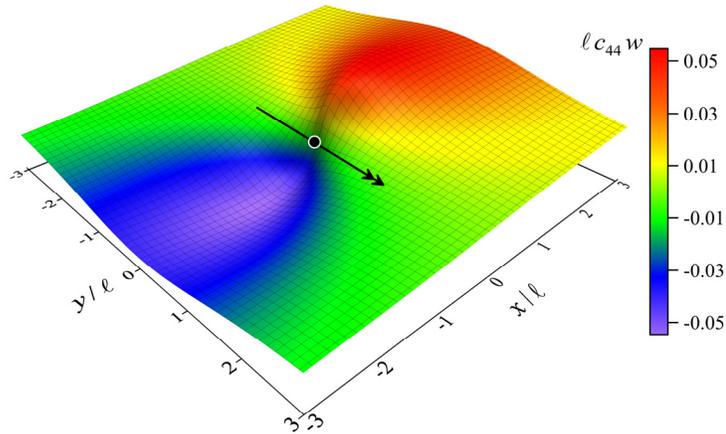

**Fig. 4**: Dimensionless out-of-plane displacement due to an antiplane concentrated unit moment (applied at the spot) in an orthotropic Cosserat solid far from material instability ( $\varepsilon = 1/4$, $\beta = 1/5$, $\gamma = 1/4$ ).

## 4. Stress channelling

It is known that for classical Cauchy elastic materials with extreme orthotropic properties the stress produced by a concentrated load has a slow diffusion, so that the solution becomes highly localized and strongly directional. In fact, in the limit when the stiffness ratio between different material directions tends to zero, the equations governing equilibrium reach the elliptic boundary and the stress percolates through *null-thickness* deformation bands. This phenomenon is called *stress channelling* and occurs in highly orthotropic fibre-reinforced materials where disturbances can propagate along (singular) fibers without attenuation (Everstine and Pipkin, 1971). Stress channelling effects have been observed also experimentally in masonry models by Bigoni and Noselli (2010a;b) and can be easily visualized in the pinscreen toy (Fig. 1, Part I).

4.1 *Classical Cauchy materials*

For a classical Cauchy material under antiplane strain conditions, stress channelling and the associated loss of (SE) and (E) occur at the limit in which the ratio between the two shear moduli tends to zero or to infinity. In particular, when $\varepsilon = c_{55}/c_{44} = 0$, (E) is lost and the shear stresses defined in Eq. (8) become in the limit (Gelfand and Shilov, 1964)

$$\sigma_{xz} = \lim_{\varepsilon \to 0} \frac{-S\sqrt{\varepsilon}\, x}{2\pi\left(x^2 + \varepsilon y^2\right)} = 0, \quad \sigma_{yz} = \lim_{\varepsilon \to 0} \frac{-S\sqrt{\varepsilon}\, y}{2\pi\left(x^2 + \varepsilon y^2\right)} = -\frac{S}{2}\mathrm{sgn}(y)\delta(x), \qquad (24)$$

showing the Dirac delta singularity. Moreover, utilizing Eq. (5) with $c_{55} = 0$, so that (E) fails, and interpreting the inversion integral in the sense of distributions, the out-of-plane displacement can be derived in the following form

$$w^{(cl)}(x,y) = \frac{S}{4\pi^2}\int_{-\infty}^{\infty}\int_{-\infty}^{\infty}\frac{1}{c_{44}k_2^2}e^{-i(k_1 x + k_2 y)}dk_1 dk_2 = -\frac{S}{2c_{44}}|y|\delta(x), \qquad (25)$$

exhibiting again the Dirac delta singularity. Therefore, at ellipticity loss, the displacement exhibits a *Dirac delta discontinuity* along the singular line $x = 0$, Eq. (25), closely resembling the behavior of the pinscreen model (Fig. 1, in Part I).

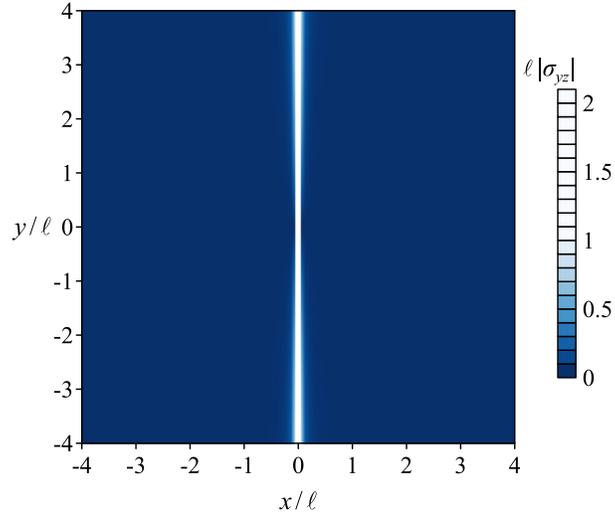

**Fig. 5**: Dimensionless modulus of shear stress due to a concentrated unit force showing that stress channelling tends to become a Dirac delta for a classical Cauchy material near ellipticity loss.

The result in Eq. (24)$_2$ reveals the extreme localization of the classical Cauchy elasticity, yielding a shear stress $\sigma_{yz}$ concentrated in a band of null-thickness when ellipticity is lost. This is substantiated in Fig. 5, where the map of the absolute value of the dimensionless shear stress $\ell\sigma_{yz}$ is reported for a classical Cauchy material close to failure of ellipticity, $\varepsilon \sim 10^{-4}$. Note that in Fig. 5, the shear stress and the coordinates $(x, y)$ are normalized with respect to the characteristic length $\ell$ of the Cosserat material, to facilitate comparisons.

4.2 *Extreme Cosserat materials*

The infinite-body Green's function for an antiplane concentrated force is used now as a perturbing agent to investigate the effects of loss of (E) in extreme, orthotropic couple-stress materials, in the spirit of the perturbative approach detailed by Bigoni (2012).

The condition of (E) for a couple-stress orthotropic material under antiplane strain deformation is recalled here to be (see also Eqs (94) and (106), in Part I)

$$b_2 n_x^4 + 2b_0 n_x^2 n_y^2 + b_4 n_y^4 \neq 0 \quad \forall \mathbf{n} : |\mathbf{n}| = 1, \tag{26}$$

which, accordingly, implies that

$$b_2 > 0 \quad \text{and} \quad b_0 > -\sqrt{b_2 b_4}, \tag{27}$$

holding for $b_4 > 0$. It is apparent that the condition of (E) does *not* depend upon the Cauchy moduli. On the other hand, the phase velocity of the dispersive SH waves in a couple-stress medium is recalled to be (see also Eq. (111), in Part I)

$$V_S^2 = \rho^{-1} \left[ c_{55} n_x^2 + c_{44} n_y^2 + \frac{k^2}{4} \left( b_2 n_x^4 + 2 b_0 n_x^2 n_y^2 + b_4 n_y^4 \right) \right]. \tag{28}$$

The (WP) condition states that $V_S^2 > 0$ for all wavenumbers $k$ and directions of propagation $\mathbf{n}$. Therefore, if the Cosserat material with $\varepsilon = 0$ is still in the elliptic range and $c_{44} \geq 0$, Eq. (28) shows that an SH wave can still propagate along *every* direction in the orthotropic medium. It is remarked that, with $\varepsilon = 0$, the underlying classical Cauchy solid (in other words the solid obtained setting the Cosserat stiffness to zero) loses ellipticity and the solution exhibits the Dirac delta behavior (Fig. 5).

Using the solution (13), the stresses have been calculated (Appendix A), so that the absolute value of the dimensionless shear stress $\ell \sigma_{yz}$ is reported in Fig. 6 for two Cosserat materials with: (i.) $\varepsilon = 0$, $\beta = 1$, $\gamma = 1$, shown on the left, and (ii.) $\varepsilon = 0$, $\gamma = 1$, $\beta = 0.01$, shown on the right. Contrary to classical Cauchy elasticity, even if $\varepsilon = 0$, stress channelling is *not* observed in the former case (Fig. 6, left), so that the Cosserat contribution restores ellipticity and *eliminates* localization. On the other hand, as the ratio of the bending moduli $\beta$ decreases, approaching the (EI/P) boundary, the stress localizes in bands of *finite* thickness and channels through the material (Fig. 6, right). Therefore, stress channelling is related to the situation that the Cosserat material approaches failure of (E). This is more clearly depicted in Fig. 7, where the shear stress profile is shown for various values of $\beta$ (left) and $\gamma$ (right), at the level $y = \ell$. It is observed that the stress localizes into a narrow band around the discontinuity line with decreasing $\beta$ and increasing $\gamma$. The width of the band depends strongly on the magnitude of the ratio of the bending moduli $\beta$.

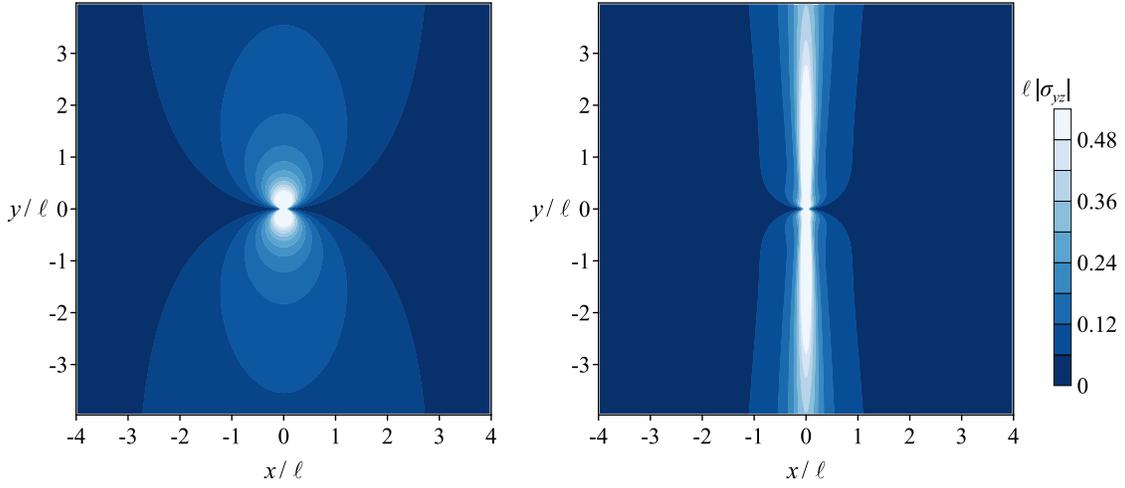

**Fig. 6**: Dimensionless modulus of shear stress due to an antiplane concentrated unit force showing that stress channelling, occurring in the underlying classical elastic material (Fig. 5), is eliminated in a Cosserat solid far from ellipticity loss, $\varepsilon = 0$, $\beta = \gamma = 1$ **(left)**, but emerges when the material is close to failure of ellipticity (EI/P), $\varepsilon = 0$, $\gamma = 1$, $\beta = 0.01$ **(right)**, so that stress channelling is related to failure of (E).

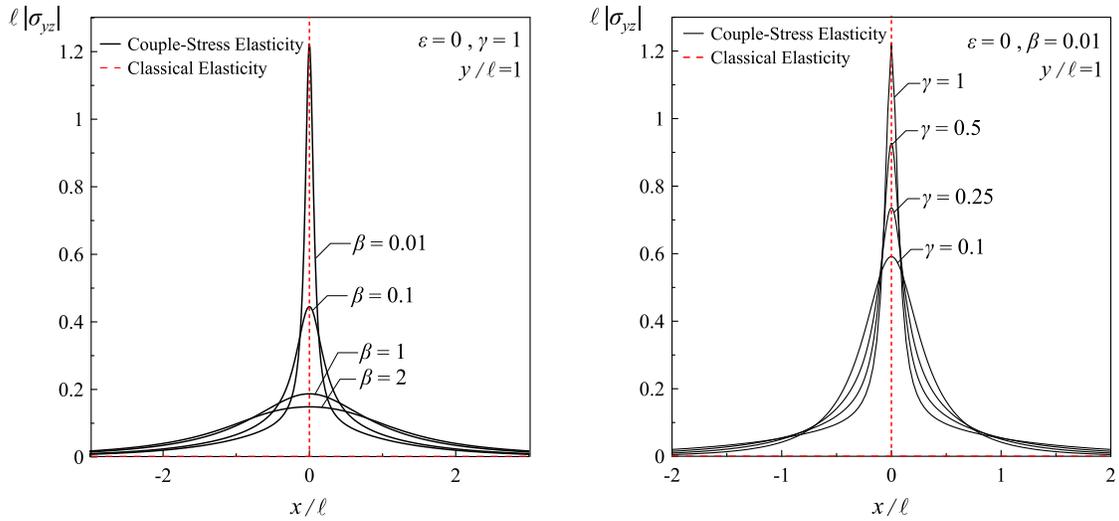

**Fig. 7**: Profiles of the dimensionless modulus of the shear stress at $y = \ell$, generated by a concentrated unit force for a material with: $\varepsilon = 0$, $\gamma = 1$, and different values of $\beta$ **(left)**; $\varepsilon = 0$, $\beta = 0.01$ and different values of $\gamma$ **(right)**. The stress in the classical elastic material localizes as a Dirac delta (solution reported red dashed), so that the addition of a Cosserat term emends this behavior until the limit of failure of (E) is approached at the (EI/P) boundary.

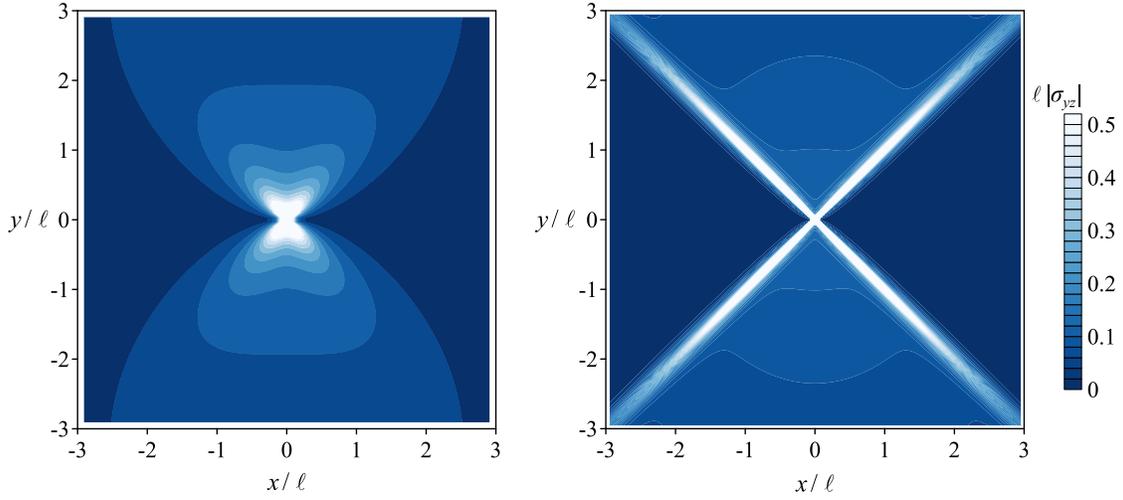

**Fig 8**: Dimensionless modulus of shear stress due to an antiplane concentrated unit force showing that stress channelling is eliminated in a Cosserat solid far from ellipticity loss $\varepsilon = \gamma = 0$, $\beta = 1$, **(left)**, but emerges when the material is close to failure of ellipticity (EC/H) $\varepsilon = 0$, $\gamma = -0.99$, $\beta = 1$, **(right)**. Note that an extreme stress channelling with *only one* localization band would be manifested in the underlying classical elastic material.

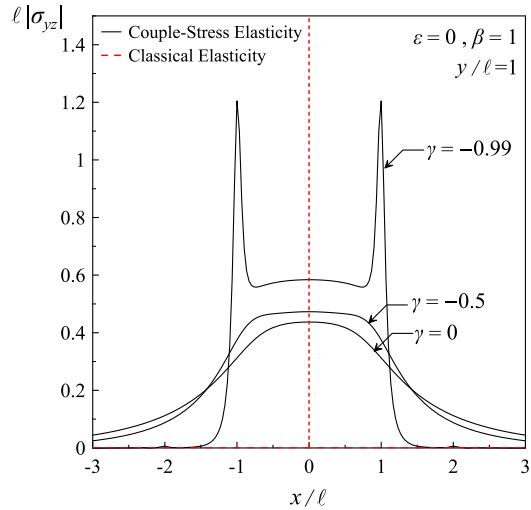

**Fig. 9**: Profiles of the dimensionless modulus of shear stress at $y = \ell$, generated by a concentrated unit force for a material with $\varepsilon = 0$, $\beta = 1$, and for different negative values of $\gamma$. The stress in the classical elastic material localizes as a Dirac delta (solution reported red dashed), so that the addition of a Cosserat term eliminates this behavior until the limit of failure of (E) is approached at the (EC/H) boundary, where two channels emerge (instead than one as in the classical theory).

A type of stress channelling *not observable in classical Cauchy elasticity* emerges as the ratio $\gamma$ takes on negative values and approaches the limit $\gamma \to -\sqrt{\beta}$, where loss of (E) is attained at the (EC/H) boundary (see also Fig. 3, Part I). Fig. 8 reports the variation of the modulus of the normalized shear stress $\sigma_{yz}$ produced by a concentrated unit force for two orthotropic Cosserat materials with: (i) $\varepsilon = 0$, $\beta = 1$, $\gamma = 0$ – far from the (EC/H) boundary – (Fig. 8, left), and (ii) $\varepsilon = 0$, $\beta = 1$, $\gamma = -0.99$ – close to the (EC/H) boundary – (Fig. 8, right). As it is expected stress channelling is *not* observed in the former case, however, as $\gamma \to -1$, approaching the (EC/H) boundary, the stress channels through two *finite width* bands *inclined* at 45° (Eq. (107), Part I). The progressive localization of the shear stress is more clearly depicted in Fig. 9, where the shear stress profile is shown for various negative values of $\gamma$, at the level $y = \ell$. Two localization bands are evidenced for the Cosserat material when failure of (E) is approached, instead than the single band typical of the classical case.

## 5. Localized single and cross folding of a Cosserat continuum

There are two cases of loss of (E) in an orthotropic couple-stress material under antiplane strain conditions. The first is at the boundary of the elliptic-imaginary/parabolic regime (EI/P), whereas the second occurs at the boundary of the elliptic-complex/hyperbolic regime (EC/H) (see also Fig. 3, Part I). It is shown below that in both these cases, the solution produced by the infinite-body antiplane Green's function (20) exhibits weak elastostatic shocks (i.e. finite jump discontinuities in certain components of the deformation gradient). On the other hand, the displacement remains continuous but displays *localized folding*, a phenomenon that cannot be captured within the context of the classical elasticity theory. Finally it should be noted that although the jumps in the components of the displacement gradient remain finite, some curvature components suffer a Dirac delta discontinuity.

*5.1 Single folding at* (EI/P) *ellipticity loss*

At the elliptic-imaginary/parabolic (EI/P) boundary, loss of ellipticity is attained when $\beta = 0$ and $\gamma > 0$ (or equivalently $b_2 = 0$ and $b_0 > 0$). *In this case, although ellipticity fails, the (WP) condition still holds (provided that $c_{44} > 0$ and $c_{55} > 0$), so that the Green's function (15) can still be obtained.*

The dimensionless displacement $wc_{44}$ depicted in Fig. 10 shows how a single, localized *folding* is formed when an antiplane concentrated unit force is applied to an orthotropic material at failure of ellipticity ($\varepsilon = 1/2$, $\beta = 0$, and $\gamma = 1$). In particular, it is observed that when (E) is lost the material folds (crumples) along the discontinuity line $x = 0$. Accordingly, the normal derivative of the displacement $\partial_x w$ exhibits a *finite* jump across the discontinuity surface, showing, thus, that the solution suffers a *weak* elastostatic shock. The magnitude of the jump depends strongly upon the parameter $\gamma$ (Fig. 11, left).

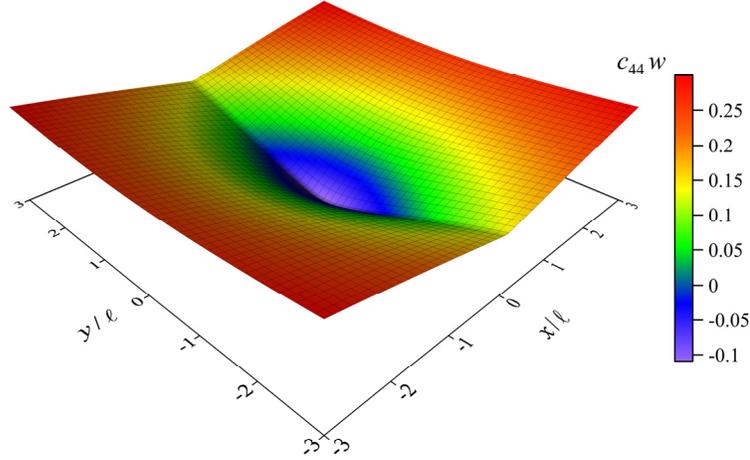

**Fig. 10**: Single folding emerging at the (EI/P) boundary for an orthotropic material ($\varepsilon = 1/2$, $\beta = 0$, and $\gamma = 1$) is evidenced by the dimensionless out-of-plane displacement produced by an antiplane concentrated unit force.

In general, at the (EI/P) boundary only one possible discontinuity surface emerges, a 'single folding', which is aligned parallel to the $y$-axis. In that case, the requirement of continuity of tractions across the discontinuity surface and the use of the Maxwell compatibility conditions (Eqs. (68)-(70), Part I), lead to the following *underdetermined* differential system for the jumps in the gradients of the out-of-plane displacement

$$\left[\!\left[ P_z^{(n)} \right]\!\right] = 0 \Rightarrow c_{55}\, g_3^{(1)} - \frac{b_2}{4} g_3^{(3)} - \frac{(2b_0+b_3)}{4}\frac{d^2 g_3^{(1)}}{dy^2} = 0, \tag{29}$$

$$\left[\!\left[ R_y^{(n)} \right]\!\right] = 0 \Rightarrow -\frac{b_2}{2} g_3^{(2)} = 0, \tag{30}$$

where $g_3^{(p)}(y) = \left[\!\left[ \partial_x^p w \right]\!\right]$ with $p = 1, 2, 3$. Note that Eqs. (29) and (30) can be directly derived from the general differential system (82) in Part I. Now, since $b_2 = 0$ at (EI/P) ellipticity loss, the second jump condition, Eq. (30), is identically satisfied, so that $g_3^{(2)}$ can be different from zero. Under these circumstances, Eq. (29) assumes the following form

$$c_{55} g_3^{(1)} - \frac{(2b_0 + b_3)}{4} \frac{d^2 g_3^{(1)}}{dy^2} = 0, \tag{31}$$

which implies that the jump in the normal displacement gradient $g_3^{(1)}$ satisfies a second-order differential equation with the following exponential decaying solution along the discontinuity line ($x = 0$)

$$g_3^{(1)}(y) = g_3^{(1)}(0) e^{-\frac{2\sqrt{c_{55}}}{\sqrt{2b_0 + b_3}}|y|}. \tag{32}$$

In fact, Eq. (32) shows that, in the case of loss ellipticity at the (EI/P) boundary, the original differential system, Eqs. (29) and (30), becomes *determinate* and therefore admits a solution, a fact substantiating the general result obtained in Part I (Section 6, Part I). It is worth noting that expression (32), describing the variation of $g_3^{(1)} = [\![\partial_x w]\!]$ along the discontinuity line, involves also the Cosserat modulus $b_3$ (i.e. the secondary bending stiffness). However, Eq. (5) shows that the determination of the infinite body Green's function does *not* involve the parameter $b_3$, but rather the parameter $b_0 = b_1 - b_3$. This implies that the discontinuity surface that is predicted to occur at failure of ellipticity can violate equilibrium, or in other words continuity of the reduced tractions (29) and (30). Nonetheless, this continuity holds if (E) and (PD)$^B$ are lost simultaneously, in which case $b_3 = 0$ and the emerging discontinuity surface becomes *admissible*.

Fig. 11 on the right, shows that the jump in the normal derivative of the displacement $g_3^{(1)}(y)$, is finite and decays *exponentially* according to Eq. (32) (with $b_3 = 0$). The magnitude of the jump depends strongly upon the parameter $\gamma$. It is observed that small values of $\gamma$ lead to jumps with large magnitude near the origin and fast decay.

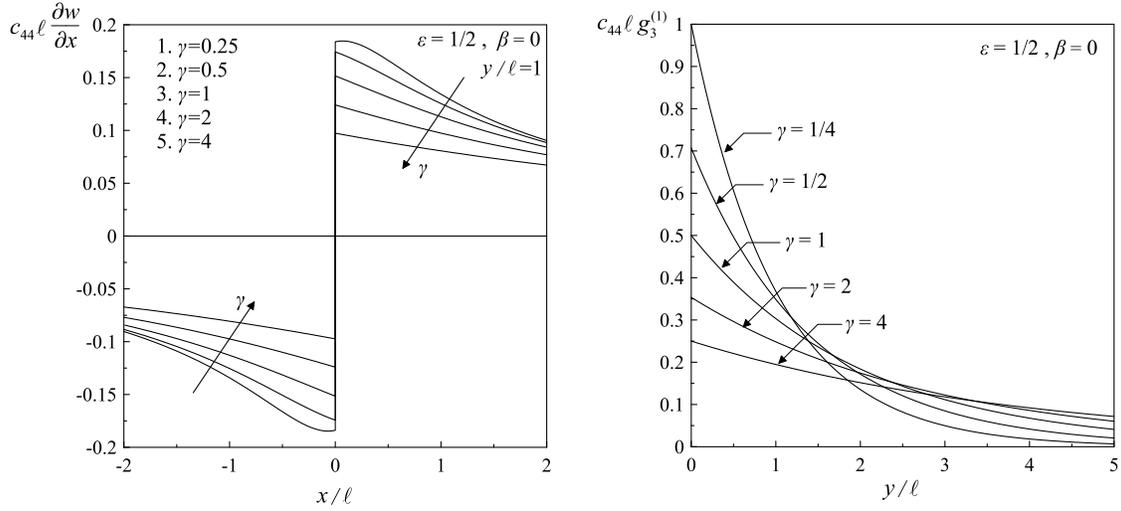

**Fig. 11**: Single folding emerging at the (EI/P) boundary as the effect of a unit antiplane concentrated force. The normal derivative of the displacement suffers a jump *across* the discontinuity line $x = 0$ **(left).** The jump $g_3^{(1)}(y)$ decays exponentially *along* the discontinuity line **(right)**.

*5.2 Cross folding at* (EC/H) *ellipticity loss*

At the elliptic-complex/hyperbolic (EC/H) boundary, loss of ellipticity is attained when $\beta > 0$ and $\gamma = -\sqrt{\beta}$ (or equivalently $b_2 > 0$ and $b_0 = -\sqrt{b_2 b_4}$). In this case, two inclined discontinuity surfaces become possible.

For instance, Fig. 12 shows how a *cross, localized folding* is formed when an antiplane concentrated unit force is applied to an orthotropic material at failure of ellipticity ($\varepsilon = 1/2$, $\beta = 1$, and $\gamma = -1$). In particular, it is observed that when (E) is lost two discontinuity surfaces are created with an inclination of $\varphi = \pm 45^o$ and a cross folding emerges along the lines $y = \pm x$. Along these folds the normal derivative of the displacement $\partial_n w$ exhibits a *finite* jump that decays away from the origin.

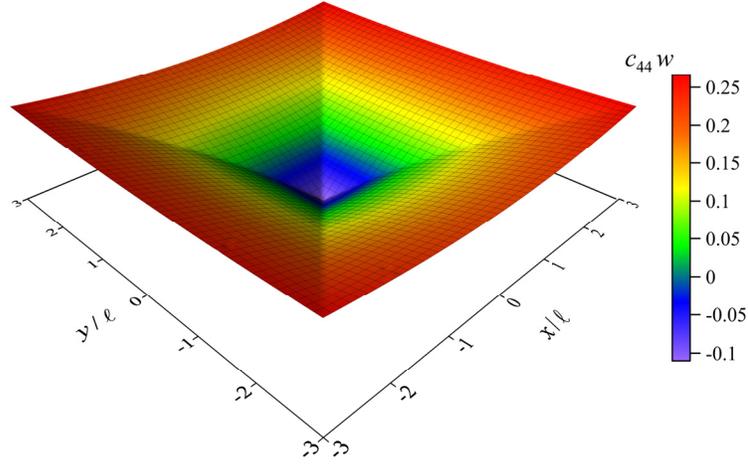

**Fig. 12**: Cross folding emerging at the (EC/H) boundary for an orthotropic material ($\varepsilon = 1/2$, $\beta = 1$, $\gamma = -1$) is evidenced by the dimensionless out-of-plane displacement produced by an antiplane concentrated unit force.

Further, employing the Maxwell compatibility conditions together with the requirement of the continuity of the tractions across the discontinuity surface, yields

$$[\![P_z^{(n)}]\!] = 0 \Rightarrow$$

$$\frac{\sqrt{b_2}c_{44} + \sqrt{b_4}c_{55}}{\sqrt{b_2} + \sqrt{b_4}} g_3^{(1)} - \left[\frac{5(b_2 + b_4 - 2b_0)\sqrt{b_2 b_4}}{4(\sqrt{b_2} + \sqrt{b_4})^2} + \frac{2b_0 + b_3}{4}\right] \frac{d^2 g_3^{(1)}}{ds^2}$$

$$+ \frac{(b_0 + \sqrt{b_2 b_4})(b_2 b_4)^{1/4}}{(\sqrt{b_2} + \sqrt{b_4})^2}\left[(\sqrt{b_2} - \sqrt{b_4})\frac{dg_3^{(2)}}{ds} - \frac{(b_2 b_4)^{1/4}}{2} g_3^{(3)}\right] = 0, \qquad (33)$$

$$[\![R_s^{(n)}]\!] = 0 \Rightarrow -(b_0 + \sqrt{b_2 b_4})(b_2 b_4)^{1/4}\left[(\sqrt{b_2} - \sqrt{b_4})\frac{dg_3^{(1)}}{ds} - (b_2 b_4)^{1/4} g_3^{(2)}\right] = 0, \qquad (34)$$

where $g_3^{(p)} = [\![\partial_n^{(p)} w]\!]$ ($p = 1, 2, 3$), and $s$ is the tangential coordinate positioned along the inclined discontinuity line. In the case of loss of ellipticity at the (EC/H) boundary ($b_0 = -\sqrt{b_2 b_4}$), the differential system described by Eqs. (33) and (34) becomes *determinate*. In fact, the condition governing the jump in the tangential moment traction, Eq. (34), is identically satisfied, whereas Eq. (33) becomes

$$\frac{\sqrt{b_2}c_{44}+\sqrt{b_4}c_{55}}{\sqrt{b_2}+\sqrt{b_4}}g_3^{(1)}-\sqrt{b_2 b_4}\frac{d^2 g_3^{(1)}}{ds^2}=0, \tag{35}$$

where it has been assumed that (E) and (PD)$^B$ are lost simultaneously, so that $b_3=\sqrt{b_2 b_4}$ and $b_1=0$ (see Eq. (90), Part I). Note that if (E) is lost *after* (PD) and *not* simultaneously, the discontinuous solution violates continuity of reduced tractions. The solution of the differential equation (35) shows that the jump in the normal derivative of the out-of-plane displacement is exponentially decaying along the discontinuity line (i.e. in the $s$-direction).

5.3 *Non-decaying single folding*

It is worth noting that in both the previously examined cases of loss of (E), SH waves still propagate in the Cosserat medium in all directions. In fact, at the (EI/P) boundary ($b_2=0$, $b_0>0$), SH waves can travel with a real non-zero velocity provided that $c_{55}>0$ and $c_{44}\geq 0$. In particular, when the direction of propagation of the SH wave is aligned with the discontinuity surface parallel to the *y*-axis, the velocity becomes: $V_S=\sqrt{c_{55}/\rho}$. On the other hand, when ellipticity is lost at the (EC/H) boundary the propagation velocity is given in the limit of $b_0=-\sqrt{b_2 b_4}$ by Eq. (28), with $b_2>0$. More specifically, when the direction of the wave propagation coincides with the direction of the discontinuity surface, the velocity of the SH waves becomes: $V_S=\sqrt{(c_{55}n_x^2+c_{44}n_y^2)/\rho}$, with $c_{55}>0$ and $c_{44}>0$, and the components of **n** are defined by Eq. (108) in Part I.

A case of interest occurs when the phase velocity of the SH waves becomes zero and *stationary* discontinuity surfaces can emerge. *Under these circumstances, the (WP) condition is also violated at the (EI/P) or (EC/H) boundaries, and a non-decaying type of folding occurs*. To investigate such a case, an extreme Cosserat material is considered with $\beta=0$ and $\varepsilon\to 0$ (i.e. $b_2=0$ and $c_{55}\to 0$), where (E) is lost at (EI/P) boundary and a single non-decaying folding emerges, parallel to the *y*-axis. It should be noted that in this case, the integral for the out-of-plane displacement in Eq. (5) *diverges* since one pair of the conjugate roots of the characteristic quartic polynomial (6) approaches zero (along the imaginary axis) for all $k_1\in\mathbb{R}$, leading, thus, to a 'near-by' double pole singularity at the origin of the real axis. However, this difficulty can be circumvented by perturbing the orthotropic body with a quasi-statically applied *dipole* of forces (two equal and opposite antiplane concentrated unit forces), placed symmetrically at a distance $2h$ along the discontinuity line $x = 0$,

$$w^d(x, y) = w(x, y - h) - w(x, y + h), \tag{36}$$

where the dipole distance is assumed to be $2h = \ell/10$.

Fig. 13 depicts the emergence of a non-decaying single folding in the case of loss of ellipticity at the (EI/P) boundary when, in addition, the (WP) condition is near to be violated ( $\varepsilon \sim 10^{-9}$ ). It is observed that the displacement $w^d$ induced by the force dipole does *not* decay in the $y$-direction. In fact, the constant displacement profile shown in Fig. 14 is obtained at every level $|y| = const$, with $|y| > 2\ell$. Moreover, the displacement produces a folding, so that the (bounded) maximum of displacement is reached on the discontinuity line $x = 0$ and decays transversally determining a displaced zone of *finite* width. Note that in Fig. 14 the displacement profiles for couple-stress and classical elasticity are compared. It is shown that the displacement for Cauchy material (dashed line) exhibits an infinite jump (Dirac delta) along the discontinuity line, so that it results localized in a zone of *null* thickness, Eq. (25), whereas the couple-stress displacement solution localizes in a band of finite thickness ruled by the parameter $\gamma$.

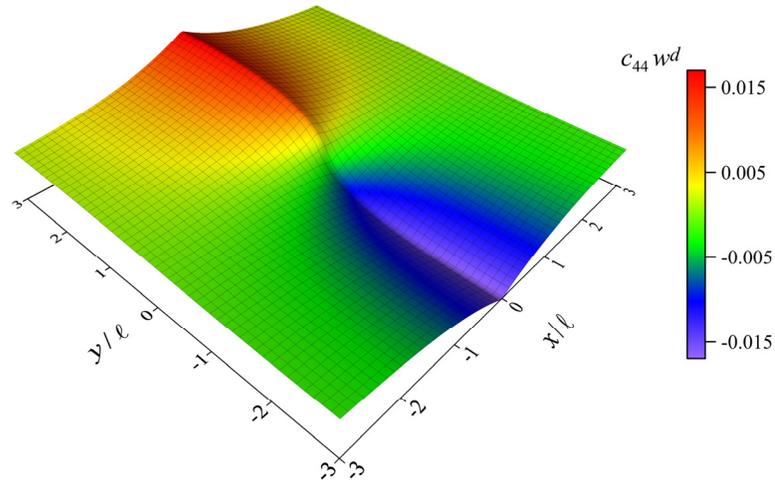

**Fig. 13**: Non-decaying single folding forming at the (EI/P) boundary for a Cosserat orthotropic material ( $\varepsilon \sim 10^{-9}$, $\beta = 0$, $\gamma = 1$ ) is evidenced by the dimensionless out-of-plane displacement $w^d$ produced by a unit force dipole (two equal and opposite forces applied at $(0, \pm h)$).

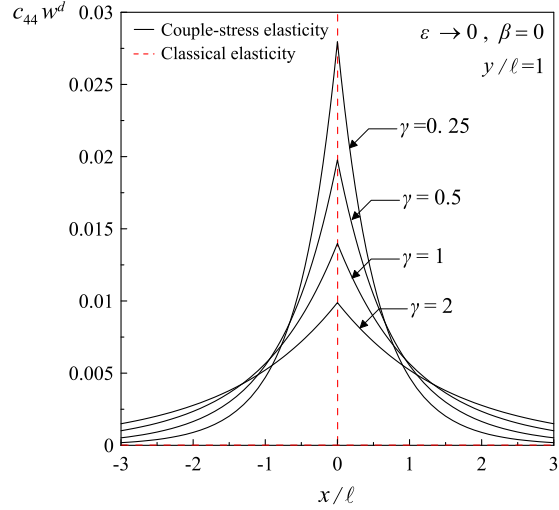

**Fig. 14**: Non-decaying single folding at the (EI/P) boundary for an orthotropic material ($\varepsilon \sim 10^{-9}$, $\beta = 0$, $\gamma = 1$) produced by a unit force dipole (two equal and opposite forces applied at $(0, \pm h)$). Comparison between the couple-stress and classical displacement profiles (the latter reported dashed and red) for different values of $\gamma$.

5.4 *An analytical solution showing decaying & non-decaying folding in two orthogonal directions*

In the extreme case of a Cosserat material with all *null* bending rigidities $b_2 = b_3 = b_4 = 0$ (so that the only non-vanishing couple-stress moduli is the torsional stiffness $b_1 > 0$), a closed form solution for the Green's function can be derived, showing simultaneously a *decaying* single folding aligned with the $x$-axis and a *non-decaying* single folding aligned with the $y$-axis. It should be remarked that in this extreme case both (E) and (WP) conditions are violated. In particular, using Eq. (5) and interpreting the inversion integral in the sense of distributions, the Green's function is obtained as

$$w(x,y) = \frac{S}{4\pi^2} \text{F.P.} \int_{-\infty}^{\infty} \int_{-\infty}^{\infty} \frac{e^{-i(k_1 x + k_2 y)}}{c_{44} k_2^2 + (b_0/2) k_1^2 k_2^2} dk_1 dk_2 = -\frac{S}{2\sqrt{2c_{44} b_0}} |y| e^{-\sqrt{\frac{2c_{44}}{b_0}}|x|}, \qquad (37)$$

with $b_0 = b_1$.

Fig. 15 depicts the variation of the normalized displacement $w^d$ induced by a dipole of out-of-plane forces for an extreme Cosserat material with $b_1 = c_{44}\ell^2$ and $c_{44} > 0$. It is observed that, immediately outside the dipole zone ($|y| > h$), the displacement profile becomes constant in the $y$-direction, so that the disturbance does *not* decay with increasing distance from the dipole. Moreover, a localization zone of finite width is observed for displacement, where the percolation thickness is a function of the material microstructure. In particular, the localization zone in Fig. 15 appears to be narrower ($\sim 6\ell$) and with a sharper profile than that reported in Fig. 14. As a conclusion, for an orthotropic Cosserat material at the (E) boundary, the presence of a torsional stiffness ($b_1 > 0$) suffices to prevent extreme localization of displacement, even when all the other couple-stress moduli are vanishing.

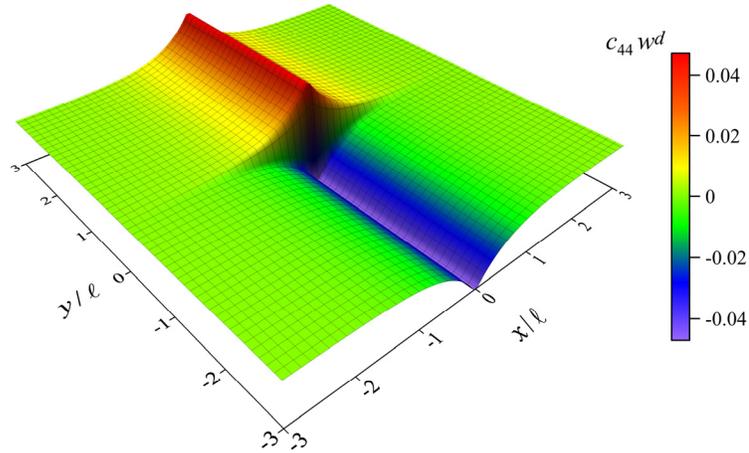

**Fig. 15**: Non-decaying single folding emerging at the (E) boundary for an extreme Cosserat orthotropic material with two non-vanishing stiffnesses (the shear stiffness $c_{44}$ and the torsional stiffness $b_0 = c_{44}\ell^2$, while $c_{55} = 0$, $b_2 = b_3 = b_4 = 0$) is evidenced by the dimensionless out-of-plane displacement $w^d$ produced by a unit force dipole (two equal and opposite forces applied at $(0, \pm h)$).

5.5 *An analytical solution showing Dirac localization*

In the case where the only non-vanishing couple-stress moduli is $b_4$ (which is the bending stiffness in the $x$-direction), the Green's function for the displacement can be derived from Eq. (5), with $c_{55} = 0$, as

$$w(x,y) = \frac{S}{4\pi^2} \text{F.P.} \int_{-\infty}^{\infty}\int_{-\infty}^{\infty} \frac{e^{-i(k_1 x + k_2 y)}}{c_{44}k_2^2 + (b_4/4)k_2^4} dk_1 dk_2 = -\frac{S\delta(x)}{4c_{44}}\left(2|y| + \sqrt{\frac{b_4}{c_{44}}} e^{-2\sqrt{\frac{c_{44}}{b_4}}|y|}\right). \quad (38)$$

Solution (38) implies that the displacement suffers an extreme localization (as in the case of classical elasticity), exhibiting a Dirac delta singularity [note, however, that the structure of the out-of-plane displacement is now different than the classical one, Eq. (25)]. This case corresponds to the origin point ($\beta = 0$ and $\gamma = 0$) in the regime classification reported in Fig. 3 of Part I. Therefore, the extreme localization, displaying a Dirac delta variation, occurs because now Cosserat effects do not improve material stability, a circumstance that can be visualized with the deck of cards example sketched in Fig. 16. In particular, this extreme Cosserat material can be visualized as a (frictionless) deck of cards that can display a bending stiffness in the plane of the cards, but this stiffness does not preclude a Dirac delta solution for the displacement, when subject to a concentrated force parallel to a card.

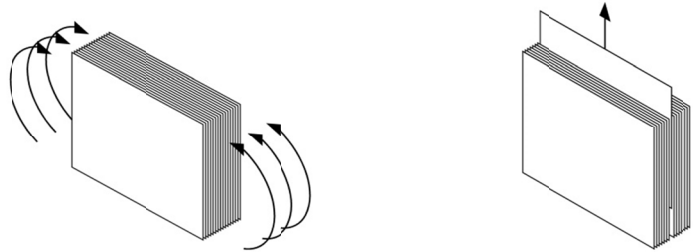

**Fig. 16**: A (frictionless) deck of cards can display a finite bending stiffness in the plane of the cards (**Left**); however, this bending stiffness does not change the response of the deck of cards to a concentrated force in the plane of the cards, which resembles a Dirac delta, as in the classical case (**Right**).

## 6. Single and cross faulting in a Cosserat continuum

It is shown in this Section that the application of a concentrated antiplane moment on an extreme Cosserat material at failure of ellipticity yields the emergence of faulting (elastostatic shocks of *finite* amplitude) in single and cross geometries. It is worth noting that there is no counterpart of such type of deformation in the classical theory of elasticity.

For the case of a concentrated unit moment (assumed parallel to the $y$-axis), the two situations of loss of ellipticity, on the (EI/P) and the (EC/H) boundary are respectively considered in Figs. 17 and 18, where the dimensionless displacement $w\ell c_{44}$ is plotted. In particular, Fig. 17 shows the formation of *single faulting* along the discontinuity line $x = 0$, at loss of ellipticity at the (EI/P) boundary for an orthotropic Cosserat material with $\varepsilon = 1/2$, $\beta = 0$, and $\gamma = 1$. On the other hand, at the (EC/H) boundary, two discontinuity surfaces emerge. For instance, assuming $\varepsilon = 1/2$, $\beta = 1$ and $\gamma = -1$, the inclination of the discontinuity surfaces becomes $\varphi = \pm 45^o$ (Fig. 17).

Finally, it should be noted that although the jumps in the out-of-plane displacement remain *finite*, so that they can be represented as Heaviside functions for both cases of single and cross faulting, the normal (to the discontinuity line) derivatives of the displacement suffer a Dirac delta type of jump.

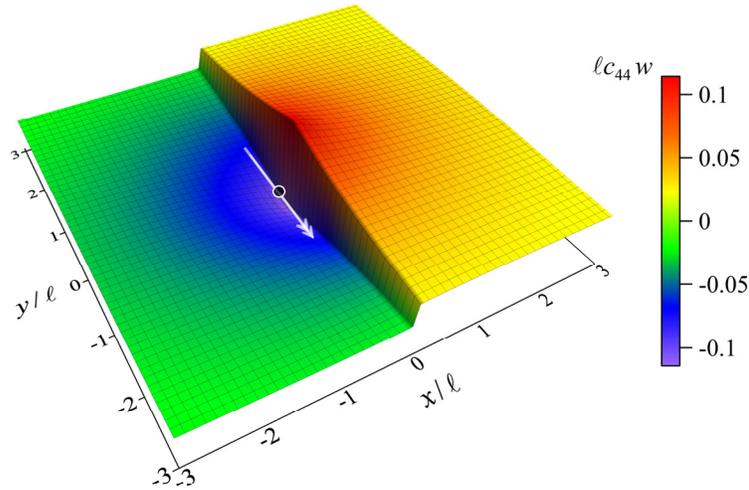

**Fig. 17**: Single faulting forming at the (EI/P) boundary for a Cosserat orthotropic material ($\varepsilon = 1/2$, $\beta = 0$, $\gamma = 1$) is evidenced by the dimensionless out-of-plane displacement produced by an antiplane concentrated unit moment (applied at the spot) parallel to the $y$-axis.

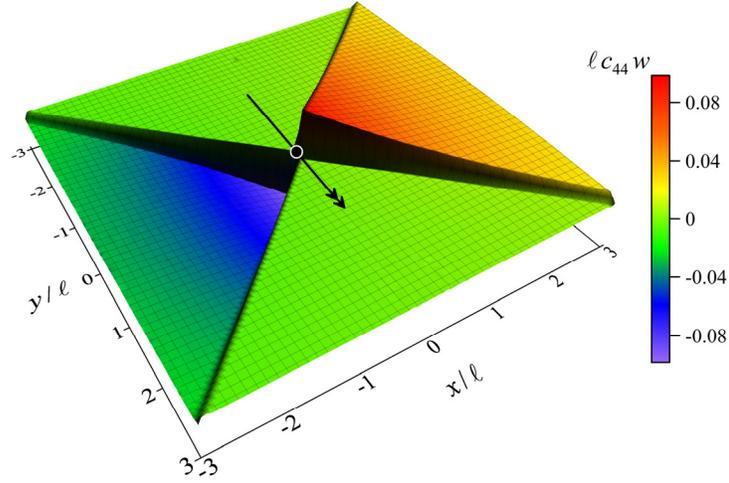

**Fig. 18**: Cross faulting forming at the (EC/H) boundary for a Cosserat orthotropic material ( $\varepsilon = 1/2$ , $\beta = 1$, $\gamma = -1$ ) is evidenced by the dimensionless out-of-plane displacement produced by an antiplane concentrated unit moment (applied at the spot) parallel to the $y$ -axis.

## 7. Conclusions

Extreme Cosserat materials have been defined in Part I of this study to be close to a material instability either for the Cauchy or the Cosserat part of the constitutive equations. For these materials uniqueness conditions (positive definiteness of the elastic energy and strong ellipticity of the elastic tensor) and stability conditions (propagation of plane waves with non-vanishing speed and ellipticity) were introduced. It is remarkable that wave propagation and ellipticity have been found to be not interdependent conditions. The former is necessary and sufficient for the determination of the infinite-body Green's function, the latter provides the condition for the formation of stress channelling, and of folding or faulting in a couple stress elastic continuum, which may occur in a single or cross modes. Note that the conclusions regarding Green's functions have been obtained under the restrictive hypothesis of antiplane strain, while the more complex case of plane strain will be addressed elsewhere.

Note that although the emergence of folding and faulting for extreme Cosserat materials has been demonstrated in the limit situation of loss of (E), these effects are already clearly visible when the material still has a strictly positive defined strain energy, but is close to failure of (E). This means that extreme materials displaying effects such as folding or faulting can be designed and realized in practice, which opens new perspectives in the design of ultra-performant materials.

*Acknowledgements* - Financial support from the ERC advanced grant 'Instabilities and nonlocal multiscale modelling of materials' FP7-PEOPLE-IDEAS-ERC-2013-AdG (2014-2019) is gratefully acknowledged.

# Appendix A. Evaluation of the shear stresses in the case of concentrated force

Upon substituting the displacement solution (13) into the constitutive equations (87) and (88) in Part I, the (asymmetric) shear stresses become

$$\sigma_{xz}(x,y) = \frac{S}{2\pi}\int_0^\infty \Sigma_{xz}(k_1,y)\sin(k_1 x)\,dk_1,\qquad\text{(A1a)}$$

$$\sigma_{yz}(x,y) = \frac{S}{2\pi}\int_0^\infty \Sigma_{yz}(k_1,y)\cos(k_1 x)\,dk_1,\qquad\text{(A1b)}$$

with

$$\Sigma_{xz}(k_1,y) = \frac{i}{2k_1}\left[p_1\left(1-2c_{44}\Delta^{-1}\right)e^{-p_1|y|} + p_2\left(1+2c_{44}\Delta^{-1}\right)e^{-p_2|y|}\right],\qquad\text{(A2a)}$$

$$\Sigma_{yz}(k_1,y) = \frac{\operatorname{sgn}(y)}{2}\left[\left(1-2c_{44}\Delta^{-1}\right)e^{-p_1|y|} + \left(1+2c_{44}\Delta^{-1}\right)e^{-p_2|y|}\right].\qquad\text{(A2b)}$$

Note that for $k_1 \to 0$ the functions $\Sigma_{xz}$ and $\Sigma_{yz}$ are bounded and, therefore, the decomposition employed for the evaluation of the out-of-plane displacement in Eq. (15) is not needed. On the other hand, as $k_1 \to \infty$ the above functions exhibit the following asymptotic behavior

$$\lim_{k_1\to\infty}\Sigma_{xz}(k_1,y) = \frac{i}{2}\left[m_1 e^{-m_1 k_1 |y|} + m_2 e^{-m_2 k_1 |y|}\right],\qquad\text{(A3a)}$$

$$\lim_{k_1\to\infty}\Sigma_{yz}(k_1,y) = \frac{\operatorname{sgn}(y)}{2}\left[e^{-m_1 k_1 |y|} + e^{-m_2 k_1 |y|}\right],\qquad\text{(A3b)}$$

with $m_{1,2} = \sqrt{\gamma \pm \sqrt{\gamma^2 - \beta}}$. Thus, utilizing the Abel-Tauber theorem (Davies, 2002), the shear stresses in the vicinity of the concentrated force behave as

$$\sigma_{xz}(x,y) = \frac{S}{4\pi} \sum_{j=1}^{2} \frac{m_j x}{x^2 + m_j^2 y^2}, \quad \sigma_{yz}(x,y) = \frac{S}{4\pi} \sum_{j=1}^{2} \frac{m_j y}{x^2 + m_j^2 y^2} \quad \text{as} \quad r \to 0. \tag{A4}$$

Consequently, the Cauchy-type singularity of the stresses is retained just as in the classical theory, although the detailed structure of these fields is altered. However, it is worth noting that although the stresses are singular, their symmetric part is not.